\documentclass[journal,10pt,onecolumn]{IEEEtran}

\usepackage[utf8]{inputenc}
\usepackage{cite}
\usepackage{amsmath,amssymb,amsfonts}
\usepackage{graphicx}
\usepackage{textcomp}
\usepackage{xcolor}
\usepackage{CJKutf8}
\usepackage{float}
\usepackage{bm}

\usepackage{geometry}
\usepackage{hyperref}
\hypersetup{
    colorlinks=true,
    linkcolor=blue,
    citecolor=blue,
    urlcolor=blue
}

\newcommand{\tao}{\begin{CJK*}{UTF8}{gbsn}韬\end{CJK*}}

\begin{document}

\title{Huawei's $\tau$ Chip Was Supposed to Melt?}

\author{Tingbo~He%
\thanks{Tingbo He is with Huawei Technologies Co., Ltd.}}

\markboth{Preprint submitted to arXiv}%
{He: Huawei's $\tau$ Chip Was Supposed to Melt?}

\maketitle

\begin{abstract}
Heat is the sharpest concern on $\tau$ scaling law --- the time-scaling principle behind Huawei's folded silicon, named for the delay $\tau$ (``Tao'') it sets out to shrink. Our Kirin measurements turned that objection upside down. The reason is one every commuter already knows: on a chip, as in a workday, it is the travel that burns the majority of the energy, not the work at the desk.
\end{abstract}

\begin{IEEEkeywords}
LogicFolding, $\tau$ scaling law, 3D integration, hybrid bonding, low-power design, system-technology co-optimization.
\end{IEEEkeywords}

Ask a hundred chip engineers what the most important technology is to the future of the industry and many will answer: 3D integration --- stacking dies on dies to boost the amount of silicon you can devote to a system. Ask any experienced chip engineer what will kill 3D integration, and you will almost always hear the same word: heat.

The reasoning feels like a law of nature. Stack active transistors on top of active transistors, cram more into every square millimeter, seal the whole thing inside a compact mobile device, and the outcome seems inevitable. Power turns into heat. Heat trapped in a confined space raises temperature. Temperature past a limit throttles the clock, degrades reliability, and eventually cooks the chip. By this logic, the denser you fold, the hotter you burn --- and the harder you fold, the faster you hit the wall.

Huawei cannot wait for that argument to settle itself. Cut off from the tools that let its competitors keep shrinking transistors, the company had to break the cycle --- and break it sooner than anyone else. So it stopped scaling space, the linear dimensions of a transistor, and started scaling time: the characteristic delay a signal takes to travel a chip's critical paths. That delay is written as $\bm{\tau}$ --- pronounced tau, and rendered in Chinese as Tao (\tao), and $\tau$ scaling law is a campaign to drive $\tau$ down generation after generation~\cite{ref1}. Its flagship technique for compressing that delay is called \textbf{LogicFolding}: taking a circuit that used to sprawl across a flat plane and folding it into vertically stacked tiers bridged by tiny vertical links so that signals travel shorter distances.

From the start, $\tau$ was conceived as a law of electronic systems rather than a device trick: time can be compressed at every level --- device, circuit, chip, and system. LogicFolding is one technique within that framework, working at the circuit and chip levels. The distinction matters, because the industry keeps forgetting it. Soon after the transistor came an invention of equal consequence --- the integrated circuit. Yet semiconductor progress is still told as a story about better transistors, as if the circuits integrated from billions of transistors were an afterthought. $\tau$ scaling inherits that blind spot: seeing the same transistors inside a $\tau$ chip, skeptics conclude that nothing fundamental has changed. The evidence that follows shows why they are wrong.

When LogicFolding was first shown in public --- in May 2026, at the IEEE International Symposium on Circuits and Systems --- the skepticism in the room was thick. The single sharpest objection pointed toward heat, heat, and more heat.

There is just one problem with that logic. Measurements on real silicon --- the Kirin 2026, set to ship this month --- showed the opposite. The chip was cooler than its predecessor while comprising 55\% more transistors per square millimeter and delivering as much as 66\% power reduction on some key tasks. What follows is why that's the case and why LogicFolding and $\tau$ scaling law will carry Huawei --- and probably others --- well into the future.

\section{What is LogicFolding?}

Moore's Law has carried the computing industry, and many others, a very long way, but it has always leaned on a single trick: geometric scaling~\cite{ref2,ref3}. Make the transistor smaller, and chips get faster and thriftier at once. More than a decade ago the trick began to wear out. Clever changes to the transistor's structure bought another decade, but every chipmaker now pays more and gets less with each new node~\cite{ref4,ref5,ref6,ref7}.

For Huawei, the squeeze came earlier and harder. Shrinking past what the industry calls the 7-nanometer node requires extreme ultraviolet (EUV) lithography, which cannot be accessed. Another way had to be found.

Six years of exploration led to a reappraisal of geometric scaling for what it always was: a means, not an end. The point of shrinking transistors was never smallness itself but the things smallness delivered --- faster switching, shorter response times, higher clock frequencies. In short, LogicFolding is a way of buying time without shrinking space. Within roughly the same silicon area, it packs in more transistors and lets the systems built from them run faster.

The enabling craft is called 3D hybrid bonding, and it is closer to welding than to packaging. Two wafers --- or a wafer and a chip --- are carefully aligned, pressed together, and slowly heated. During the anneal, covalent bonds form across the two oxide surfaces, merging into a single dielectric layer, while the copper pads on each face expand toward each other and fuse into continuous metal. The result is a forest of vertical connections between the stacked dies. In some of the most advanced commercial logic chips today, those connections sit about 10~$\mu$m apart, or fewer than 1 million vertical links per 100~mm$^2$.

The real design question is how many vertical connections there are per unit area before folding actually cuts RC --- the resistance and capacitance that slow and tax every signal. There is no closed-form answer, but one piece of intuition is reliable. A typical logic process has a top-metal pitch of 720~nm. Match the bonding pitch to that figure, and face-to-face wafers can join top-metal signals directly. Choose a sparser pitch, and those signals must fan out to sparser pads --- adding wire, congestion, and RC, the very costs folding is meant to remove. Design-space exploration confirmed the intuition: 1.5~$\mu$m already supports good LogicFolding, 1~$\mu$m is better, and 720~nm would expand the benefit further still. In the future, we expect to shrink the bonding pitch further, to 480~nm, so hybrid-bonding pads can connect to lower metal layers and give LogicFolding more design freedom.

We treat the hybrid bonding as a wafer-level, cross-layer device step rather than a packaging step. Die-to-wafer bonding is a die-handling process --- pick and place --- evolved from the chip-packaging factory. Its accuracy baseline is low, and even after years of tightening it still sits at the micrometer level. Wafer-to-wafer bonding starts from a different world: wafer-fab handling, where lithography tools already hold overlay to few nanometers. That high-accuracy baseline is the right starting point, but bonding still has to solve problems the litho scanner never faced --- stress control, CMP planarization, cleaning, wafer bow, edge taper, and alignment --- before it can deliver sub-micrometer pitch. On 40-nm tooling we reached a 1.5~$\mu$m hybrid-bonding pitch in Kirin 2026 --- 50 million vertical interconnects, of which 10 to 15 percent carry signals. Kirin 2027 silicon is already at 1~$\mu$m and more than 100 million vertical interconnects. Within three years we expect to match the 720~nm top-metal pitch, a one-to-one gear ratio, and more than 200 million vertical interconnects.

With the vertical links in place, the folding could begin. Huawei's engineers examined the hardest-working blocks of the Kirin SoC --- the NPU, the CPU, the GPU and the DSP --- and sought out the longest, slowest, most power-hungry horizontal routes inside each one. Each block was then redesigned to occupy two tiers of silicon at once, turning those long, horizontal connections into short vertical hops from one die to another.

On Kirin 2026, the first mobile system-on-chip (SoC) relying on LogicFolding, transistor density rose in a single generation from roughly 155 to 238 million transistors per square millimeter --- a 55\% jump. That's an equivalent increase to what we achieved in the previous three years using geometric shrinking.

A great result of course, but intuition says the power density should have climbed in lockstep.

It fell.

Compared to the planar predecessor, at matched performance, Kirin 2026 achieves power reduction of 66\% on the NPU, 58\% on the GPU, and 41\% on the CPU performance core at the real-world benchmark test. We put substantially more transistors into every square millimeter, and the chip ran cooler per unit of work, not hotter.

That result deserves an explanation, because it looks like a violation of the very physics the skeptics invoked. It is not. It is a lesson about where a chip's power actually goes --- and the answer is not where most people assume.

\section{A Tale of Two Power Bills}

In a smartphone's typical daily workload, its chips expend power in two ways. One, static leakage, or the power lost even when no computing is happening, is minimal. Its opposite, dynamic power, what's lost during computing, makes up roughly 90\% of the total power bill.

Digital circuits essentially consume dynamic energy in two ways. The first is \textbf{switching}: flipping a transistor gate, charging and discharging the tiny capacitance of a logic gate to change a bit from 0 to 1 and back. This is the ``real work'' --- the computation itself.

The second is \textbf{moving signals}: charging and discharging the long metal wires that carry those bits from one part of the chip to another. A wire is a capacitor smeared out over distance. The farther a signal travels, the more capacitance you charge, and the energy scales with that capacitance. In the textbook expression for dynamic power,
\begin{equation}
P = \alpha \cdot C \cdot V^{2} \cdot f
\label{eq:dyn_power}
\end{equation}
the term $C$ (capacitance) is not dominated by the transistors. On an advanced node, it is dominated by wires. Interconnect capacitance, not gate capacitance, is where the energy goes.

Here is what trips up the intuition: in a modern SoC block, the wires --- the moving --- routinely consume more energy than the gates --- the computing. The processor spends most of its power not thinking but commuting.

\section{How Your Commute Explains Your Chip}

Consider an average office worker's daily energy budget. It is tempting to assume most of it is spent on the actual job --- the meetings, the spreadsheets, the thinking. It isn't. For many office workers, the largest energy expenditure of the day is the \textbf{commute}: the drive or the train ride between home and office, twice a day, every day. Desk work is comparatively cheap; transport dominates.

A chip is no different. The gates do the thinking and the wires do the commuting. It is just as true in a large AI cluster: more than 80\% of the energy is spent moving data, not computing on it. The same imbalance, in milder form, lives inside every smartphone SoC.

Once you see power this way, the thermal objection to folding inverts. Critics assumed packing transistors closer must raise power density, because they implicitly focused on counting gates --- the desk work. But the dominant term is the commute. And what does LogicFolding do? It moves everyone's home closer to the office.

When you fold a circuit into vertical tiers connected by hybrid bonding pads at sub-micrometer pitch, signals that once crossed long horizontal wires (hundreds of micrometers) now drop straight down through a bonding pad a few micrometers away. In our experience, wire lengths on folded paths fall by 20\% on typical cores and by up to 70\% on some critical paths. Clock networks --- some of the busiest, most power-hungry wires --- shrank so much that we cut clock-buffer count by more than half. On one processing block alone, folding shortened the clock wiring by 28\%, and dropped the buffer count from 43,600 to 19,000.

Every shortened wire is a smaller capacitance to charge, so the dominant term in the power equation collapses. The desk work --- the actual switching --- is essentially unchanged, but the commute has been slashed. That is where the power-density reduction comes from. Not from computing less, but from moving data less.

\section{Kirin 2026 Power Consumption, Block by Block}

The blocks dominating a smartphone SoC's thermal budget are precisely those moving the most data: the NPU running AI inference, the GPU shading pixels, and the DSP crunching signals. These run hot for the same reason: they are communication-intensive. Because they shuttle enormous volumes of data between memory, buffers, and arithmetic units, a large fraction of their power is commute, not compute --- and that is exactly the fraction that folding attacks.

To show how much each block gained from LogicFolding, we measured it by running the folded chip against its planar predecessor, the Kirin9030 Pro, \textbf{at iso-performance}: same AI throughput, frame rate, or benchmark score. A block that needs less speed to do the same work can run at a lower voltage, and the savings compound. Meanwhile, as the same silicon can also run flat out, each block is reported in both modes: matched to its predecessor, and \textbf{at full throttle}. Here is what the silicon reported for the blocks.

\begin{figure}[H]
\centering
\includegraphics[width=0.8\linewidth]{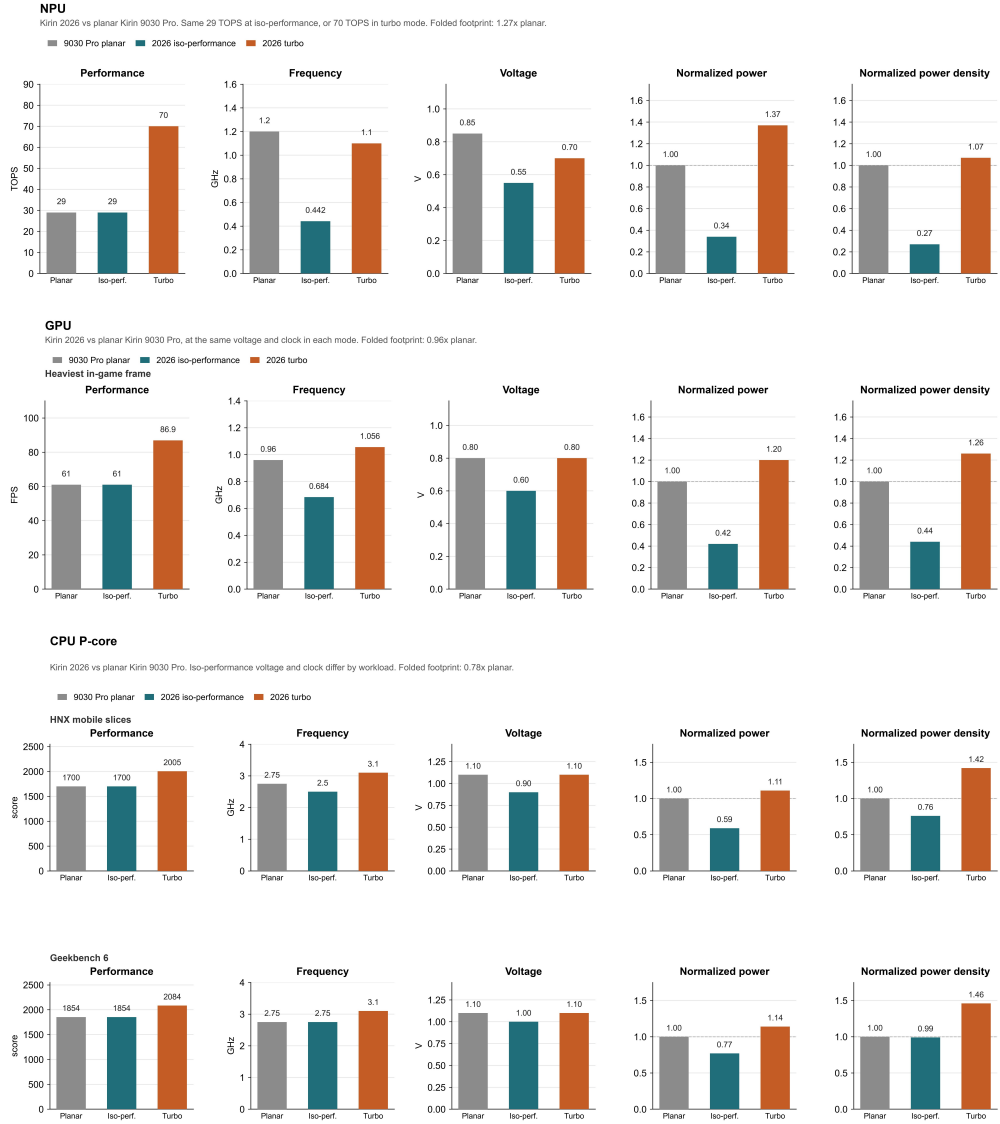}
\caption{Performance, frequency, voltage, normalized power, and normalized power density of the planar Kirin9030 Pro (gray), the Kirin 2026 at iso-performance (teal), and the Kirin 2026 in turbo mode (orange). The GPU's heaviest workload is the most intensive rendering frame from a specific game; CPU HNX bench comprises key mobile application slices, measuring real-world CPU performance and power; Geekbench 6 is an industry-standard CPU performance benchmark. HNX focuses on the real-world user experience, whereas Geekbench 6 targets the heaviest performance workloads.}
\label{fig:block}
\end{figure}

As shown in Fig.~\ref{fig:block}, each chart has five panes --- performance, frequency, voltage, normalized power, and normalized power density --- and three bars: the planar Kirin9030 Pro, the Kirin 2026 at iso-performance, and the Kirin 2026 in turbo mode. The dashed line on the normalized panes marks the 9030 Pro baseline. Folded footprints are the projected area of the two-tier stack --- the area that power density is measured against.

\begin{figure}[H]
\centering
\includegraphics[width=0.8\linewidth]{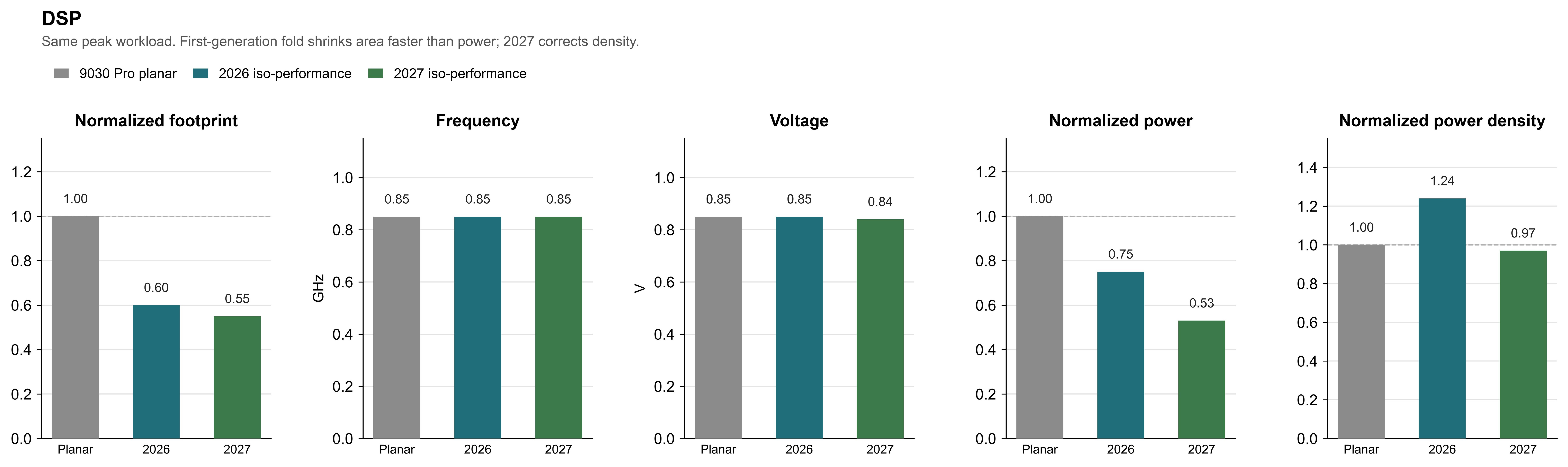}
\caption{Normalized footprint, frequency, voltage, normalized power, and normalized power density of the planar Kirin9030 Pro (gray), the Kirin 2026 at iso-performance (teal), and the Kirin 2027 at iso-performance (green).}
\label{fig:dsp}
\end{figure}

The NPU is the block critics most fear stacking because it is both the densest and among the hottest. Yet it posts the largest reduction of all: at the same 29 TOPS, its clock frequency could be dialed down by 63\% and its voltage could be reduced from 0.85~V to 0.55~V. That leads to a reduction in power of 66\%, and a decline in power density of a staggering 73\%. The GPU is close behind: operating at the same 61 FPS as Kirin9030 Pro but at a voltage that's 200~mV lower, power fell by 58\%. The gradient runs exactly along the commute: the more parallel and data-hungry a block, the more folding pays it back.

The DSP is the exception, and it teaches an important lesson: with LogicFolding, area can shrink faster than power (Fig.~\ref{fig:dsp}). The first-generation fold (Kirin 2026) does the same work with 25\% less power, but power density --- what's consumed per square millimeter --- rose 24\%, and it is power density that really matters to controlling heat. The reason for the increase is straightforward: folding shrank the DSP's footprint by 40\%, so its power density climbed even as its power consumption fell. The second-generation fold (Kirin 2027) settled the matter: we already have that silicon and have measured a lower power density than the planar original along with a 47\% drop in power consumption. Density, in other words, is a design output, not a fate.

The CPU performance core gains handsomely from folding --- matching its predecessor's benchmark score at a 9\% lower clock and 200~mV less supply voltage, consuming 41\% less power on the HNX benchmark test. But its power-density saving is the most modest of the hot blocks. That is not a failure of folding; it is a clue to something important we'll talk about in a moment.

Of course, we aren't only interested in power savings. We also ran the Kirin 2026 blocks flat out to see what they could do compared to the 9030 Pro. Run at full throttle, the folded NPU delivers 70 TOPS --- 141\% more than its predecessor, the GPU delivers 42\% more frames, and the CPU delivers an 18\% boost on the HNX benchmark score; in those modes their power densities rise above the planar predecessor chip's. So folding hands the system a control dial that the planar chip never had: the same block can be a miser or a record-setter depending on the scenario.

\section{Parallel Theories: $\tau$ Scaling Law in Circuits, Amdahl's Law in Software}

Shortening the commute explains some of the power savings, but even more came from reducing voltage, because dynamic power is proportional to the square of voltage. LogicFolding unlocks this reduction by sorting circuits into three kinds: the \textbf{truly serial}, demanding peak frequency and voltage; the \textbf{parallelizable circuits}, spreading work across many copies running side by side; and those circuits lying in between, partly serial and partly parallel~\cite{ref8}.

The uncompromisingly serial circuits are the minority. Overwhelmingly, a modern SoC's transistors serve the second and third categories, the NPU, GPU, and DSP, which draw an outsized amount of power precisely because they are wide, parallel, and data-hungry. The CPU performance core (big core) is the clearest inhabitant of the first category, because its single fast integer thread of execution cannot be parallelized, and it must run near the top of the voltage and frequency curve. Recall that folding buys it only a modest 9\% clock reduction at iso-performance on HNX benchmark --- the linear commute saving, not the full quadratic voltage one. Applying LogicFolding to low-parallelism CPU performance core requires meticulous design, advanced EDA toolchains, and long verification cycles. It is a systemic overhaul, not a quick fix. While substantial gains have been delivered in Kirin 2026, the bulk of the work lies ahead, requiring aggressive, exploratory innovation over the next three to five years. Yet, on the path to these ultimate yields, tangible and compounding improvements will be realized every single year.

Fortunately, these architectural hurdles do not have to be overcome by silicon alone. Smartphones have something that a naive analysis misses: deep hardware-software co-design. Rather than forcing one heroic big core to shoulder every demanding workload, task schedulers and folded floorplans are co-optimized to parallelize tasks in software and dispatch them across several small, efficient cores. This keeps the serial fraction of the work that the big core must own as small as the application allows. This shifts workloads from category one to category two, where the quadratic voltage lever applies.

Here, both levers combine. Folding shortens wires while freeing up room for many more transistors within the same footprint. Those extra transistors can be spent creating many copies of circuits which can then be run in parallel, spreading out the work so each can run at slower clock and lower voltage. Where the commute argument buys a linear saving in $C$, the parallelism argument buys a quadratic saving in $V^{2}$.

The folded NPU is a prime example. While its predecessor comprised one big core and two efficiency cores, the transistor surplus from folding paid for four big cores. The wider array delivers 70 TOPS at 0.7~V --- well below the 0.85~V its predecessor needed to deliver less than half that. More hardware, running gentler, doing more: the quadratic lever, working exactly as the equation dictates.

That is why a denser chip drew less power. The serial work requiring high voltage is small enough not to dominate the power budget; everything else --- the vast parallel majority --- is folded wider, clocked gentler, and run at lower voltage.

This design mirrors Gene Amdahl's famous insight: a program's serial fraction limits how much parallelism can buy you. Amdahl's law is about software. What LogicFolding exposes is its \textbf{hardware twin}: the serial fraction of the silicon is small, so most of the chip can trade speed for parallelism. They stand as twin stars --- one governing algorithms in code, the other governing delay and energy in circuit, forming the theoretical bedrock of System-Technology Co-Optimization (STCO).

\section{Designing Away Hot Spots}

Despite all the power savings we've accomplished with LogicFolding, one could still argue that we'll have a heat problem. Stacking two tiers of busily-switching transistors concentrates heat, because what's generated at the bottom tier must move through material of the top tier to escape.

We contend that critics are aiming at the wrong metric. What matters is not the heat, but the \textbf{heat density} --- and, one level deeper, the \textbf{junction temperature} of transistors, which is the metric that actually governs speed, leakage, and reliability. A chip fails not from dissipating a certain number of watts, but when a transistor junction climbs past its operating temperature limit. Every thermal argument worth making is ultimately an argument about that junction temperature.

Two things keep it in control. First, folding lowers heat density at the source: total power genuinely dropped at iso-performance and the reduced power is distributed across multiple active silicon tiers rather than a single densely populated plane. This achieves lower watts-per-cubic-millimeter, and heat density is the thing that sets how hot a junction gets.

Second, the remaining heat is managed. Because folding shrinks the footprint of the NPU and other hot blocks, there is room to allow the heat to spread out horizontally, eliminating localized hotspots and reducing junction temperatures, before it travels vertically to the surface. The modest vertical temperature gradient where the bottom tier runs a few degrees warmer than the top, is budgeted as a design parameter by placing the hottest blocks where heat escapes most easily. The genuine enemy is not average temperature, but the localized \textbf{hotspot} where junction temperature spikes far above its neighbors. Folding earns its thermal headroom when the design tools actively interleave hot and cool logic across tiers, preventing hotspots from stacking directly atop one another.

That is why our Kirin 2026 implementation was deliberately conservative: folding applied selectively along the critical paths that gain most, with thermal-aware placement rather than blindly stacking everything. Done that way --- lowering heat density, spreading horizontally, budgeting vertical gradients, and hunting down hotspots --- the junction temperature stays inside its window.

\section{Sisyphus, but the Hill Leads Somewhere}

It would be dishonest to end on a note of final victory.

$\tau$ is a time scaling law, not an energy law. A folded system running faster but burning more power violates no time-scaling principle --- yet it would drain your battery by lunchtime. The cool-running Kirin happened because we deliberately spent the time headroom of folding on lower voltage and power. That trade is a companion discipline, not a free consequence of the law.

The folded NPU doing its predecessor's work on a third of the power can also outrun that predecessor by 141\%, exceeding its power density. Physics permits both settings; discipline chooses the cool one. It's a choice we plan to make with each new generation of chips.

The folded CPU is deliberately conservative, targeting only critical paths. Yet, it already restores performance-core frequencies to 3.1~GHz this year. Unlocking the full potential of LogicFolding over the next three to five years requires aggressive, industry-wide innovation --- expanding the design space, shrinking hybrid bonding pitch, integrating lower-metal TSVs, vertically stacking more tiers, et al. LogicFolding will pave the way for Kirin CPU core frequency to reach 5~GHz and beyond. The roadmap is feasible and economically viable.

In the myth, Sisyphus is condemned to push his boulder up the hill, only to watch it roll down and start over. Engineering under $\tau$ scaling law shares this rhythm: fold the chip, earn the time headroom, spend it on power --- then the next generation arrives and the boulder starts over. The above proposed folded wiring is not an escape from that cycle; it is the next push. Where our story parts ways with the myth is that this hill leads somewhere: every cycle leaves behind a chip that computes more and burns less, and the accumulated distance is what the industry calls progress. The work is never finished --- that is not the law's weakness, it is the job description. And one must imagine the engineers happy.

\section{Cooler Is the Way Forward}

But heat --- the objection everyone reaches for first --- turns out to rest on a misunderstanding of where a chip spends energy. The fear counts the desk work. The physics is dominated by the commute. LogicFolding shortens the commute, lowering time constant $\tau$ and energy in one stroke, because distance costs both.

Every future we can see for $\tau$ is won or lost on power. Time headroom is the instrument; energy is the prize. The next decade of this law will be judged not by how fast the folded stacks run but by how little they burn while running.

The $\tau$ chip that was supposed to melt ran cooler instead --- not in spite of the folding, but because of it. $\tau$ (Tao), in the end, cuts in more directions than one.

\end{document}